\documentclass{ws-procs9x6}

\begin{document}

\title{Renormalization of Singular Potentials and Power Counting}

\author{M. Pav\'{o}n Valderrama$^*$ and E. Ruiz Arriola}

\address{Departmento de F\'{i}sica At\'{o}mica, Molecular y Nuclear, 
Universidad de Granada,\\
Granada, E-18071, Spain\\
$^*$E-mail: mpavon@ugr.es}

\begin{abstract}
We analyze the renormalization of the nucleon--nucleon interaction at
low energies in coordinate space for both one and two pion exchange
chiral potentials. The singularity structure of the long range
potential and the requirement of orthogonality respectively
determines, once renormalizability is imposed, the minimum and maximum
number of counterterms allowed in the effective description of the
nucleon--nucleon interaction in a non-perturbative context.
\end{abstract}

\keywords{Renormalization; NN interaction; Two Pion Exchange.}

\bodymatter

\section*{}
\vspace{-1.0cm}

The nucleon--nucleon (NN) interaction can be better understood if we take into
account that there is a separation of scales between long and short range 
physics, which is the basis of the Effective Field Theory (EFT) formulation 
of nuclear forces~\cite{Bedaque:2002mn}.
The long range piece of the interaction, $V_L$, is given by pion exchanges, 
and its form is unambiguously determined by the imposition of chiral symmetry,
while the short range piece $V_S$ is a zero-range potential 
(i.e. $V_S(r) = 0$ for $r > 0$) which represents the NN contact terms.
This last piece is regularization-dependent and determines the number 
of free parameters, or counterterms, of the EFT description.

\begin{figure}[ttt]
\begin{center}
\epsfig{figure=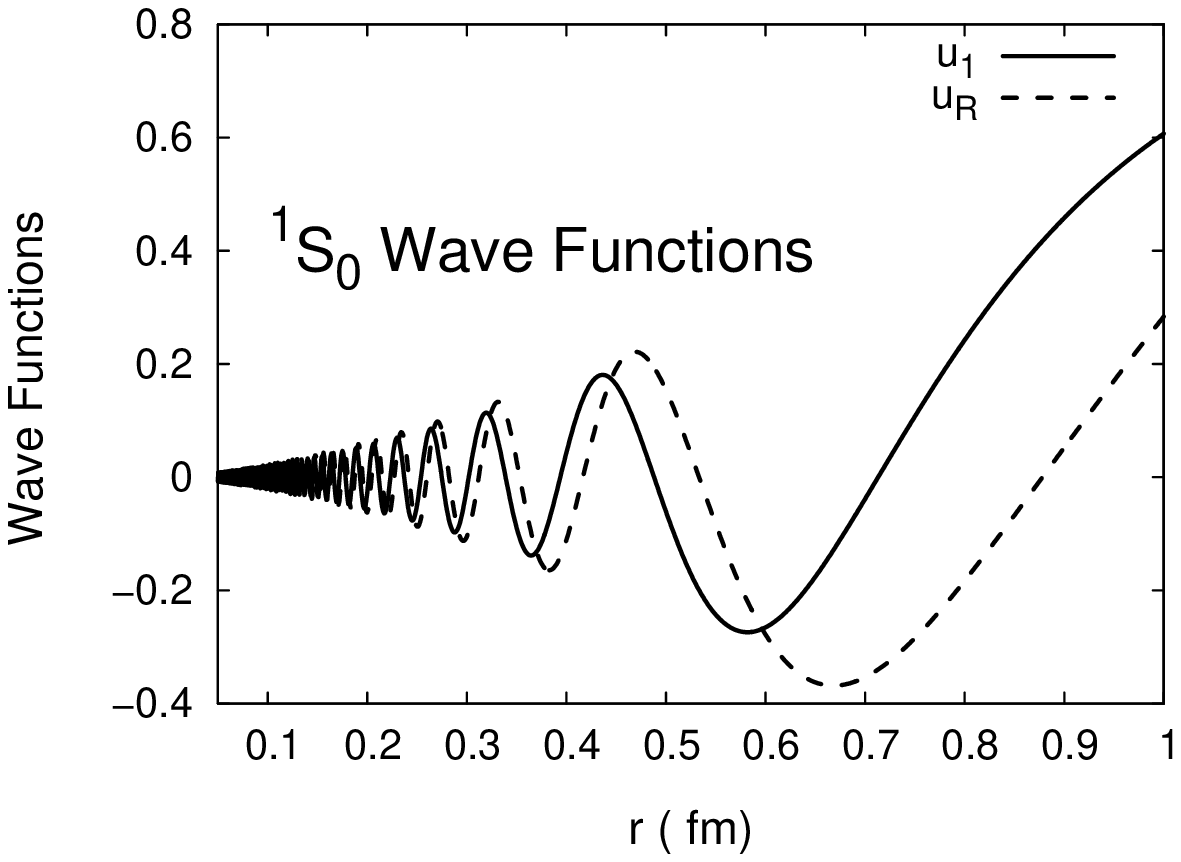,
	height=4.0cm, width=5.6cm}
\epsfig{figure=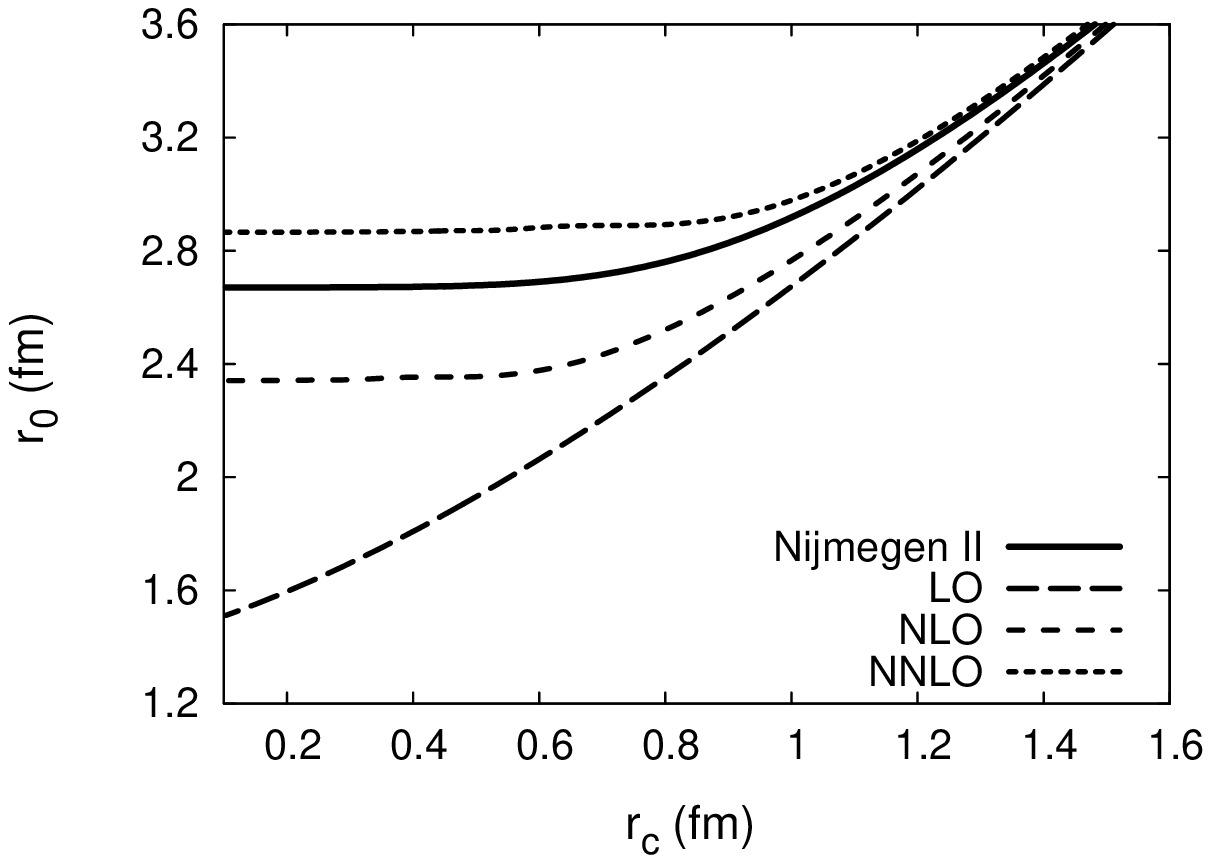,
	height=4.0cm, width=5.6cm}
\end{center}
\caption{
(Left panel) Two zero-energy, linearly independent wave functions for the 
$^1S_0$ singlet channel at NNLO; $u_1$ and $u_r$ respectively behave 
as $1$ and $r$ for $r \to \infty$. 
(Right Panel) Value of the effective range $r_0$ as a function to the cut-off 
for the same channel and different orders; $r_0$ is computed by $r_0(r_c) = 2\,[\int_0^{\infty}(1-r/\alpha_0)^2
\,dr -\int_{r_c}^{\infty}u_0^2\,dr]$, with $\alpha_0 = -23.74 {\rm fm}$ 
the scattering length.
\vspace{-0.35cm}
}
\label{fig:s-wave}
\end{figure}

In Weinberg's power counting these potentials can be written as an expansion 
in terms of the dimensionality of the contributions,
$V_{NN} = V_{LO} + V_{NLO} + \dots\,$, where $V_{LO}$, $V_{NLO}$, 
and so on, represent increasing order contributions to the NN potential.
Standard EFT wisdom states that the power countings of the long and short 
distance potentials are independent, regardless of the renormalization 
procedure, although there is a lively discussion on the field about this 
issue~\cite{Nogga:2005hy,Epelbaum:2006pt,Birse:2005um,PavonValderrama:2005gu,Valderrama:2005wv},
and, as it is shown below, if one takes the long distance potential seriously 
from $r > 0$ to infinity, this cannot be the case. 
One easy example is provided by the $^1S_0$ neutron--proton scattering 
state at NNLO, which can be described by the s-wave reduced Schr\"odinger 
equation, i.e. $-u'' + M_N V_L(r)\,u(r) = k^2\,u(r)$, with $V_L$ the long
range potential at NNLO, which is attractive,
and displays a $-1/r^6$ singularity near the origin. 
As a second order differential equation, the Schr\"odinger equation
has two linearly independent solutions, so how does one choose the physical 
solution? The regularity condition $u(0) = 0$, equivalent to the assumption 
that there is no short distance physics, cannot be applied for this case,
since any solution $u(r)$ is regular at the 
origin~\cite{PavonValderrama:2005gu,Valderrama:2005wv}, 
as shown in Fig.~(\ref{fig:s-wave}).
Then we are forced to use a boundary condition near the origin, equivalent
to adding a counterterm to the theory.

And how many more counterterms can be added? According Weinberg's power 
counting, there are two counterterms for the $^1S_0$ wave at NNLO. 
But if we take into account orthogonality between different energy solutions, 
we find out that the boundary condition must be energy independent, 
meaning that one can only have one 
counterterm~\cite{PavonValderrama:2005gu,Valderrama:2005wv} 
(any extra counterterm breaking orthogonality).
So one is led to this alternative when removing the cut-off. 
The prediction of the effective range for the singlet, 
$r_0 = 2.86\,{\rm fm}$ (the experimental value is 
$2.77 \pm 0.05\,{\rm fm}$), can serve as an orthogonality test.

\bibliographystyle{ws-procs9x6}

\end{document}